# Modelling Art Interpretation and Meaning.

# A Data Model for Describing Iconology and Iconography

SOFIA BARONCINI, MARILENA DAQUINO, FRANCESCA TOMASI[*]

**ABSTRACT:** Iconology is a branch of art history that investigates the meaning of artworks in relation to their social and cultural background. Nowadays, several interdisciplinary research fields leverage theoretical frameworks close to iconology to pursue quantitative Art History with data science methods and Semantic Web technologies. However, while Iconographic studies have been recently addressed in ontologies, a complete description of aspects relevant to iconological studies is still missing. In this article we present a preliminary study on eleven case studies selected from the literature and we envision new terms for extending existing ontologies. We validate new terms according to a common evaluation method and we discuss our results in the light of the opportunities that such an extended ontology would arise in the community of Digital Art History.

## 1. Introduction[1]

Iconology is a branch of art history that investigates the meaning of artworks in relation to their social and cultural background. The field, initiated by Aby Warburg's studies[2], evolved into a multidisciplinary approach leveraging sociology and the history of culture to read artworks as witnesses of a social memory[3]. Nowadays, several interdisciplinary research fields leverage theoretical frameworks close to iconology to pursue Digital Art History, that is, the study of Art History through the lenses of quantitative methods[4]. Communities of scholars, archivists, and museum representatives have largely investigated how to represent the Arts domain through formal languages, such as ontologies[5], and new datasets are constantly produced[6]. In particular, Semantic web technologies have been increasingly recognized as powerful means to serve and leverage machine-readable cultural heritage data in innovative applications[7].

However, while iconographic studies have been addressed in ontologies[8], a complete description of aspects relevant to iconological studies is still missing, such as the identification of cultural phenomena, symbolic parallelisms in a visual work. Representing such aspects is challenging, since heterogeneous interpretive situations populate the field of Art interpretation, and it is not always possible to explain high level, vague, or uncertain, relations in a deductive process.

---

[*] Sofia Baroncini, sofia.baroncini4@unibo.it, Digital Humanities Advanced Research Centre (/DH.arc), University of Bologna; Marilena Daquino, marilena.daquino2@unibo.it, Digital Humanities Advanced Research Centre (/DH.arc), University of Bologna; Francesca Tomasi, francesca.tomasi@unibo.it, Digital Humanities Advanced Research Centre (/DH.arc), University of Bologna.

[1] S. Baroncini is responsible for Sections 3, 4 and 5. M. Daquino is responsible for section 2 and 6. F. Tomasi is responsible for section 1 and 7.
[2] A. WARBURG, *The renewal of pagan antiquity. Contributions to the cultural history of the European Renaissance*, Getty Research Institute for the History of Art and the Humanities, Los Angeles 1999.
[3] O. R. PINELLI, *La Storia delle storie dell'arte*, Einaudi, Torino 2014.
[4] L. MANOVICH, *Data science and digital art history*, in «International Journal for Digital Art History», vol. I, 2015.
[5] M. DOERR, *The CIDOC conceptual reference module: an ontological approach to semantic interoperability of metadata, in* «AI magazine», vol. XXIV, n. 3, 2003, p.75. M. DOERR, S. GRADMANN, S. HENNICKE et. al., *The europeana data model (edm).,* in «World Library and Information Congress: 76th IFLA general conference and assembly», vol. X, August 2010, p. 15. V. A. CARRIERO, A. GANGEMI, M. MANCINELLI et. al., *ArCo: The Italian cultural heritage knowledge graph*, in *International Semantic Web Conference*, edited by Springer, Cham, October 2019, pp. 36-52. M. DAQUINO, F. MAMBELLI, S. PERONI et. al., *Enhancing semantic expressivity in the cultural heritage domain: exposing the Zeri Photo Archive as Linked Open Data*, in «Journal on Computing and Cultural Heritage (JOCCH)», vol. X, n. 4, 2017, pp. 1-21.
[6] C. A. KNOBLOCK, P. SZEKELY, E. FINK et. al., *Lessons learned in building linked data for the American art collaborative*, *International Semantic Web Conference*, edited by Springer, Cham, October 2017, pp.263-279. C. DIJKSHOORN, L. JONGMA, L. AROYO *et. al., The Rijksmuseum collection as linked data*, in «Semantic Web», vol. 9, n.2, 2018, pp. 221-230.
[7] V. ALEXIEV, *Museum linked open data: Ontologies, datasets, projects*, In «Digital Presentation and Preservation of Cultural and Scientific Heritage», vol. VIII, 2018, pp. 19-50.
[8] N. CARBONI, L. DE LUCA, *An Ontological Approach to the Description of Visual and Iconographical Representations,* in «Heritage», vol. II, n. 2, 2019, pp. 1191–1210, doi: https://doi.org/10.3390/heritage2020078. R. GARTNER, *Towards an ontology-based iconography*, in «Digital Scholarship in the Humanities», vol. XXXV, n. 1, pp. 43-53.

Nonetheless, being able to represent such aspects as machine-readable data according to shareable models would be the cornerstone of a new branch of Digital Art History dedicated to Iconology analytics. Benefits include (1) the creation of appropriate data sources suitable for Machine Learning tasks (e.g., artwork classification) or reasoning tasks, (2) the discovery of relations between artefacts preserved in different institutions, (3) the identification of patterns between cultural phenomena and motifs, symbols, and subjects.

In this article we survey theoretical frameworks for defining Art interpretation and we present the application of a representative one in a few case studies. The objective is to address most interpretative situations emerged from the literature in a bottom-up approach, and to provide a data model to guide new projects in quantitative iconological studies. We formally represent scenarios by reusing terms belonging to existing ontologies as much as possible, and we propose new terms only when existing ones are not sufficient to answer Competency Questions (CQ) relevant to iconological studies. In the light of good practices in ontology reuse[9], we do not propose another ontology, but we envision terms as specializations or extensions of existing ones.

The article is organised as follows. In section 2 we review existing vocabularies and ontologies for iconography and iconology, highlighting the gaps. In section 3 we survey existing theoretical frameworks on Art interpretation. In section 4 we explain the approach of our work, and in section 5 we present case studies to exemplify our method and the usage of the data model. In section 6 we propose a preliminary evaluation, and we discuss results.

## 2. Related work

The classification of artworks according to their iconographic apparatus has been addressed in several notable works. Iconclass, is a classification system for iconographical description with a standardized, controlled terminology, widely used for research purposes and information retrieval in museums[10]. Similarly, the Getty Iconography Authority[11] includes proper names, relationships, and dates for iconographical narratives, characters, events, literary works. Despite being excellent solutions for supporting most iconographic recognition tasks, there are naturally situations that cannot be represented. For instance, subjects cannot be related as representations of the same meaning (see section 5.1). Moreover, a subject can present different motifs over time according to the style of the period (see section 5.2). Getty Vocabularies allow to associate a date to a subject, but the stylistic variants cannot be related to the respective historical period. Likewise, Iconclass and Getty Vocabularies allow to specify the source of the iconography, but it is not possible to relate a source to a variant of a subject.

Complex relations between subjects can be better expressed with ontologies, which provide the appropriate semantics. CIDOC-CRM[12] is the ISO standard proposed by the ICOM consortium to formally represent museum data. The rationale of the model stands behind common methodologies for tracing the history of collections of cultural objects. The ontology allows to represent subjects as conceptual or visual entities and to loosely relate them. To tackle aspects related to iconography, the Visual Representation Ontology (VIR)[13] has been recently proposed. VIR (prefix vir:) extends CIDOC-CRM with specialistic terminology relevant to iconographic recognition and allows to address the role of visual cues (such as attributes, personifications, characters), their participation to a broader representation, and their symbolic value. Still, no terminology is dedicated to the relation between meanings and real-world phenomena that may influence the recognition of subjects.

---

[9] V. CARRIERO et al., *The landscape of ontology reuse approaches*, in «Applications and Practices in Ontology Design, Extraction, and Reasoning», vol. XLIX, n. 21, 2020.
[10] <http://iconclass.org/help/lod> (last consulted: 21/04/2021).
[11] <http://153.10.241.9/research/tools/vocabularies/cona/about.html#ia> (last consulted: 21/04/2021).
[12] M. DOERR, *op. cit.*, p.75.
[13] N. CARBONI, L. DE LUCA, *op. cit.*

In this study, we reuse CIDOC-CRM and VIR as much as possible to cover aspects relevant to artwork description and iconographic interpretations. However, reasons, methods, relations between claims, and sources used to support an iconological recognition cannot be fully represented. For this reason, we selected the SPAR Ontologies[14] for describing the bibliographic domain, and HiCO[15], an ontology for representing hermeneutical aspects, to integrate prior ontologies.

## 3. Theoretical background

Aby Warburg's studies[16] are unanimously recognized as the first attempt to develop a multidisciplinary approach leveraging sociology and the history of culture to read artworks as witnesses of a social memory[17], that is, iconology. The term iconology used to be often confused with iconography. The aim of iconography is to study attributes (e.g. depicted objects) and changes in the representation of subjects (e.g., identifying the representation of *Time* as winged and with a sickle in Renaissance, and depicted as Kairos in classical iconography[18]). Iconology, instead, focuses on the interpretation of iconographic subjects as documents of socio-cultural phenomena - therefore explaining the reasons of iconographical changes. In this work we rely on the definition of iconology given by Warburg and Panofsky, despite we acknowledge other general theories exist[19].

When defining interpretive aspects characterising iconology, several theories have been proposed, each defining a different arrangement of levels of interpretation that contribute to the analysis of an artwork. A level of interpretation can be defined as the grouping of claims made by an art historian with respect to an observable phenomenon, such as the recognition of depicted attributes and characters, claims on the symbolic meaning of attributes, or claims on the socio-cultural implication of a subject. In Table 1 we summarize alternative definitions of interpretation levels as provided by some of the most notable art historians, namely: Panofsky[20], Van Straten[21], Wittkower[22], Gombrich[23], Imdahl[24].

|        | Panofsky | Van Straten | Wittkower | Gombrich | Imdahl | |
|--------|----------|-------------|-----------|----------|--------|---|
| Lev. 1 | Pre-iconographical description | Pre-iconographical description | Literal representational level | Identification of a "representation" | Pre-iconographical description | Iconic sense of the image |
| Lev. 2 | Iconographical analysis | Iconographical description | Literal thematic level | Subject identification after recognising the genre of the artwork | Iconographical analysis | |
| Lev. 3 | Iconological interpretation | Iconographical interpretation | Multiple meaning | Acceptable meanings for the genre recognised | Iconological interpretation | |
| Lev. 4 |  | Iconological interpretation | Expressive meaning | Pluralism of meanings depending on the adopted approach | | |

Table 1: Overview of theories on the levels of artwork interpretation

The theoretical frameworks slightly disagree on the definition of layers. All the scholars agree on the definition of the first level as the description of recognizable basic shapes and objects depicted in a visual representation[25], and on the second level as the identification of iconographical subjects.

---

[14] S. PERONI, D. SHOTTON, *The SPAR ontologies*, in *International Semantic Web Conference*, edited by Springer, Cham, October 2018, pp. 119-136.
[15] M. DAQUINO, F. TOMASI, *Historical Context Ontology (HiCO): a conceptual model for describing context information of cultural heritage objects*, in *Research Conference on Metadata and Semantics Research*, edited by Springer, Cham, September 2015, pp. 424-436.
[16] Cfr. A. WARBURG, *op. cit.*
[17] O. R. PINELLI, *op. cit,* pp. 272-273
[18] E. PANOFSKY, *Studies in iconology: Humanistic themes in the art of the Renaissance*, Harper & Row, New York 1962, pp. 71-73.
[19] Cfr. W. J. T. MITCHELL, *Iconology: image, text, ideology*, The University of Chicago Press, Chicago 2013.
[20] E. PANOFSKY, *op. cit,* pp. 3-17.
[21] R. VAN STRATEN, *An Introduction to Iconography: Symbols, Allusions and Meaning in the Visual Arts*, Taylor & Francis, New York 2000, pp.3-24.
[22] R. WITTKOWER, *Allegory and the migration of symbols,* Thames and Hudson, London 1987, pp. 173-187.
[23] E. H. GOMBRICH, *Symbolic images*, Phaidon, London 1972, pp. 1-22.
[24] M. IMDAHL, *Iconica. L'intuizione delle immagini*, in «Aisthesis. Pratiche, Linguaggi e Saperi dell'estetico», vol. v, n. 2, 2012, pp. 11-32. doi: https://doi.org/10.13128/Aisthesis-11474.
[25] E. PANOFSKY, *op. cit.;* R. VAN STRATEN, *op. cit.*; R. WITTKOWER, *op. cit.*; E. H. GOMBRICH, *op. cit.*

Panofsky describes the third level as the broader iconological interpretation, while Van Straten and Wittkower split the level into a preliminary level, addressing secondary intentional meanings added by the artist, and a second including unintentional meaning. While Van Straten describes the latter as the identification of influences or signs of contemporary culture, Wittkower describes it as the investigation of the expression of the artist's character. Gombrich addresses the first two levels only and gives instructions on how to discover and validate other meanings. Lastly, Imahl adds another interpretation level concerning the iconic sense given by formal aspects of the representation.

To date, a shareable research method or a framework to support Art interpretation has not been validated yet. In addition, scholars have pointed out that the subjective intuition of the researcher is relevant, if not predominant, in the interpretive process[26].

In this study we investigate how Van Straten's theoretical framework applies to selected, representative, case studies to ground our work on a trusty (despite arguable) theory. Van Straten's work has been chosen because of the importance he attributes to the socio-cultural background of the artwork and the possibility to distinguish meanings traditionally attributed to a symbol from those depending on the unintentional contemporary socio-cultural influence. As an example, the subject *Lucrezia's suicide* (lev. 2) traditionally symbolizes moral virtue (lev. 3). Warburg[27] attributed a different meaning in the analysis of the Sassetti Chapel frescoes by Domenico Ghirlandaio. In this work, clients belonging to the Medici's family are characters of a sacred scene, which was an unprecedented example of the Florentine habit of offering portraits as votive gifts (lev. 4). To this extent, the traditional meaning of Lucrezia's suicide (lev. 3) gains a different connotation (lev. 4). We believe that being able to formally express such extraordinary situations is fundamental for pursuing data-driven digital art history, hereby facilitating quantitative analysis of diachronic changes of iconographical and iconological studies.

## 4. Research methodology and approach

The aim of this research is to formally represent the results of art historians' epistemological process and the underlying levels of interpretation by means of ontologies. We define a conceptual framework that is applicable to diverse scenarios and that can be used as a reference guide when producing machine-readable data about iconographical and iconological interpretations.

To address the variety of interpretive situations that can emerge in the artwork analysis, we created a corpus of case studies taken from the literature, wherein artworks have been analysed by notable art historians. The selection is based on the analysis of around 50 iconological studies performed by Warburg[28] and the major representatives of iconological studies, including Panofsky[29], Panofsky and Saxl[30], Gombrich[31], and Wittkower[32]. Other studies were taken into account to select case studies, namely: Cardini, Acidini Luchinat & Ricciardi[33]; Christiansen[34]; Hirn[35]; Lavin[36]; Robb[37]; Van Straten[38]. We noticed some recurring structures in the analysis pursued by the historians. We therefore grouped case studies according to shared elements and we identified four typologies, representing the following common interpretive situations:
- Attribution of subjects to different interpretation levels. In this group, relations between subjects located at different interpretive levels are analysed, spanning from the general interpretation of a representation to the study of symbols.
- Diachronic changes in a subject representation. In this group, the iconographical analysis addresses relations between different representations of the same subject. Different visual cues can be read as evidence of the culture in which the artwork was produced, such as the reception of classical themes and the evolution of a subject's significance.
- Connections with literary sources. In this group, the recognition of subjects is informed by the (alleged) artist's visual or literary sources. Different situations are here considered, namely: recognitions depending on unknown sources; illuminated manuscripts, in which miniatures contradict the text they illustrate; recognition of literary subjects and related meaning.
- Different subjects represented by the same shapes and motifs. In this group, similar shapes or motifs appear in artworks from different historical periods or contexts of production[39]. In some cases, it is possible to identify a prototype or model.

---

[26] M. G. MÜLLER, *Iconography and Iconology as a Visual Method and Approach*, in E. MARGOLIS, L., PAUWELS (ed.) *The SAGE Handbook of Visual Research Methods*, SAGE Publications Ltd, 2011, pp. 283–97.
[27] A. WARBURG, *op. cit.*
[28] *Ibidem.*
[29] E. PANOFSKY, *op. cit.*; ID., *Meaning in the visual arts: papers in and on Art History,* Doubleday, Garden City, N.Y 1955.; ID., *The iconography of Correggio's camera di San Paolo,* The Warburg Institute University of London, London 1961.
[30] E. PANOFSKY, F. Saxl, *Classical Mythology in Medieval Art*, in «Metropolitan Museum Studies», vol. IV, n. 2, 1933, pp. 228-280.
[31] E. H. GOMBRICH, *op. cit.*
[32] R. WITTKOWER, *op. cit.* ID., *Chance, Time and Virtue*, in «Journal of the Warburg Institute», vol. I, n. 4, 1938, pp. 313-321. doi:10.2307/749998.
[33] F. CARDINI, C. ACIDINI LUCHINAT, L. RICCIARDI, *I re magi di Benozzo a palazzo Medici*. Mandragora, Firenze 2001.
[34] K. CHRISTIANSEN, *Lorenzo Lotto and the Tradition of Epithalamic Painting*, in «Apollo», n.124, 1986, pp. 166–73.
K. CHRISTIANSEN, *Lorenzo Lotto / Venus and Cupid*, in «The Metropolitan Museum of Art», 2018, URL https://www.metmuseum.org/art/collection/search/436918 (last consulted: 19/03/2021).
[35] Y. HIRN, *The Sacred Shrine: A study of the poetry and art of the catholic church*, Beacon Press, Boston 1957, pp. 296-98.
[36] I. LAVIN, *Giambologna's Neptune at the Crossroads*, in I. LAVIN, *Past-present: essays on historicism in art from Donatello to Picasso*, University of California Press, Berkeley 1993, pp. 63-84.
[37] D. M. ROBB, *The iconography of the Annunciation in the Fourteenth and Fifteenth Centuries*, in «The Art Bulletin», vol. XVIII, n. 4, 1936, pp. 480-526.
[38] R. VAN STRATEN, *op. cit.*

For each typology, we selected two to three representative case studies (including one or more artworks), and we created a corpus of 11 total case studies. Each artwork has been described by one or more historians, and their analysis has been framed into four interpretation levels, namely: pre-iconographical objects are described in lev. 1; subjects are identified in lev. 2; possible meanings are proposed in lev. 3 and lev. 4. Descriptions also include relations to other artworks, literary sources, and the context of the artwork production. The formal description of each case study includes: (1) a description of the scenario in natural language, (2) the Competency Questions (CQ), (3) a graphical representation of the ontology terms, and (4) the description of usage of new terms.

It's worth noting that some limitations affect the study. Although a difference between the third and fourth level of interpretation may exist, it is not always possible to clearly distinguish the levels. We rely on interpretations made by authoritative historians, although we acknowledge the possibility that new case studies may add unexpected situations that are not currently investigated. Secondly, the framework here considered slightly differs from Van Straten's theory. While Van Straten's objective is to understand the epistemological process underlying the analysis of an artwork, this study focuses on the formal representation of results of such a process. Therefore, if a socio-cultural phenomenon (lev. 4) allows us to recognize the subject of a painting, the subject will not be located at the fourth level but at the second one.

## 5. The case studies and the ontology

In Table 2 we present the case studies, including a brief description and the problem to be addressed.

| Artwork | Case study |
|---|---|
| **Typology 1** | |
| Vermeer, *Woman Holding a Balance*, ca. 1664, Washington, National Gallery of Art | The relation between symbols in the scene allows to interpret the subject as an allegory of Divine Justice (lev. 2)[40]. |
| Lorenzo Lotto, *Venus and Cupid*, ca. 1520, New York, Metropolitan Museum | Christiansen[41] interpreted symbols as an epithalamic message (lev. 3), and the artwork as evidence of the tradition of donating paintings as wedding gifts (lev. 4). |

---

[39] E. PANOFSKY, *Studies in Iconology, cit.;* A. WARBURG, *op. cit.,* pp. 18-31.
[40] R. VAN STRATEN, *op. cit*., pp.3-24.
[41] K. CHRISTIANSEN, *Lorenzo Lotto and the Tradition of Epithalamic Painting, cit.*

| | |
|---|---|
| Benozzo Gozzoli, *Procession of the Magi*, 1459-1462, Florence, Palazzo Medici | Cardini et. al.[42], recognized the portrait of Lorenzo de Medici (lev. 2) as evidence of clients' attempt to celebrate the family (lev. 3) and a cultural habit (lev. 4). |
| **Typology 2** | |
| Parigi, Bibliothèque Nationale de France, ms 6362, *Histoire Universelle*, XV sec. Anonymous, *The Rape of Proserpina* | Panofsky[43] (1962) noticed that classic themes were transmitted during the Middle Ages by a textual tradition, untied from their classical visual appearance. These two artworks (an XV sec. miniature and a Baroque sculpture) are just one of the numerous examples of this phenomenon. Stylistic differences are associated with artistic movements or cultural contexts and are meaningful evidence of an iconological evolution (lev. 4). |
| Gian Lorenzo Bernini, *The Rape of Proserpina*, 1622, Rome, Galleria Borghese | |
| Vatican City, Biblioteca Apostolica Vaticana, *Aeneis*, XIV sec., ms vat. lat. 2761, folio 15r. Anonymous, *Laocoön and His Sons* | *Laocoön and His Sons* is compared to the classical sculpture representing the same theme. The former belongs to the tradition of classical contents represented deprived of the original classical style, wherein a stylistic difference between the medieval and the classical artworks depicting the same theme documents an iconological evolution[44]. |
| Agesander, Athenodoros and Polydorus, *Laocoön and His Sons*, 40-20 a.C., Vatican City, Vatican Museums | |
| Jan Van Eyck, *Annunciation*, 1434, Washington, National Gallery of Art | Variants of the same subject are related to different beliefs (lev. 4) on how the Immaculate Conception took place[45]. The iconological recognition depends on the comparison of elements belonging to different interpretation levels. |
| C. XV sculptor, *Annunciation*, Würzburg, Marienkapelle, gable of the north portal | |
| **Typology 3** | |
| Correggio, *Pan*, 1518, Parma, Camera di San Paolo | Panofsky[46] interpreted an unusual figure (lev. 2) by means of the reconstruction of the client's social and cultural background and the sources known in the area. The overall third level interpretation of the cycle documents style and preferences of the society (lev. 4). |
| Giambologna and Tommaso Laureti, *Neptune Fountain*, 1563-66, Bologna | The *Quos Ego* iconography is recalled by the *Neptune Fountain* to emphasize the political message (lev. 3) already existing in the original context of Virgil's *Aeneid*[47]. The original symbolic meaning in the literary source is crucial for the interpretation of the artwork's meaning (lev. 3), which has been reshaped in the new social context of the city of Bologna ruled by the papal vicelegate. |
| Marcantonio Raimondi, *Quos Ego*, 1515-6 | |
| Parigi, Bibl. Nationale de France, Marco Polo, *Livre des merveilles*, XV sec, ms 2810. Folio 85r, *Kingdom of Eli*; f. 88r, *Roc*; folio 55v., *crocodile*; folio 24r, *salamander* | The text of the manuscript contradicts the tradition on oriental lands while the illumination, in spite of the surrounding text, adheres to this tradition[48]. The contrast with the source is meaningful and leads to an iconological interpretation (lev. 4). |
| **Typology 4** | |
| *Hercules and the Erymanthian Boar*, c. III A.D., Venice, St. Mark's Basilica, | Classical visual shapes are transmitted during the Middle ages deprived of their classical content[49]. The shape represents a different subject or symbol. |
| *Allegory of salvation*, c. XIII A.D., Venice, Basilica di San Marco, external wall | |
| Gian Lorenzo Bernini, *The Rape of Proserpina*, 1622, Rome, Galleria Borghese | Bernini's *Rape of Proserpina* is the propotype of the *Rape of Cybele*, later re-interpreted as *Time and opportunity*[50]. The copy of compositional motifs between the last two subjects is motivated by similarities between attribute and figures. |
| Thomas Regnaudin, *Rape of Cybele*, 1678, Paris, Jardins des Tuileries | |
| David Le Marchand, *Time and Opportunity*, 700-1720 ca, London, V&A Museum | |

Table 2: Artworks, case studies descriptions, and typologies

---

[42] F. CARDINI, C. ACIDINI LUCHINAT, L. RICCIARDI, *op. cit.*
[43] E. PANOFSKY, *Studies in Iconology, cit.,* pp. 18-31
[44] E. PANOFSKY, F. Saxl, *op. cit.,* p. 253
[45] Y. HIRN, *op. cit.,* pp. 296-98.; D. M. ROBB, *op. cit.,* pp. 480-526.
[46] E. PANOFSKY, *The iconography of Correggio's camera di San Paolo, cit.*
[47] I. LAVIN, *op. cit.,* pp. 63-84.
[48] R. WITTKOWER, *Allegory and the Migration of Symbols*, *cit*., pp. 75-92.
[49] E. PANOFSKY, F. Saxl, *op. cit.,* p. 228
[50] R. WITTKOWER, *Chance, Time and Virtue, cit.* pp. 315-16

For the sake of brevity, we discuss only one representative case study for each typology (Fig. 1). A complete description of all case studies is available at <https://sofibar.github.io/Icon/>.

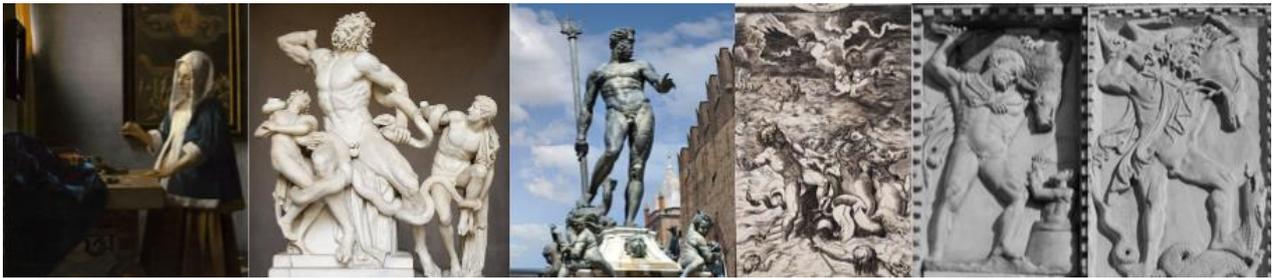

Fig. 1.a: Vermeer, *Woman Holding a Balance*, c. 1664, Washington, National Gallery of Art. ; 1.b: Agesander, Athenodoros and Polydorus, *Laocoön and His Sons*, 40-20 a.C., Vatican City, Vatican Museums ; 1.c: Giambologna and Tommaso Laureti, *Neptune Fountain*, 1563-66, Bologna; Marcantonio Raimondi, *Quos Ego*, 1515-6, detail ; 1.d: *Hercules and the Erymanthian Boar*, c. III A.D., Venice, St. Mark's Basilica, external wall; *Allegory of salvation*, c. XIII A.D., Venice, Basilica di San Marco, external wall.[51]

As aforementioned, we reuse terms from CIDOC-CRM[52] (prefix crm:) and VIR Ontology[53] (prefix vir:) whenever applicable to describe the artwork and iconographical interpretations. We reuse PRO[54] (prefix pro:) for describing roles held by characters represented in visual works. We reuse HiCO[55] (prefix hico:) and CiTO[56] (prefix cito:) to represent historians' interpretations and their usage of sources. New terms are specified with the prefix *icon*.

In Fig. 2 we show an overview of the terms relevant to iconological analysis grouped by the interpretive levels as defined by Van Straten. New classes and properties are highlighted in red.

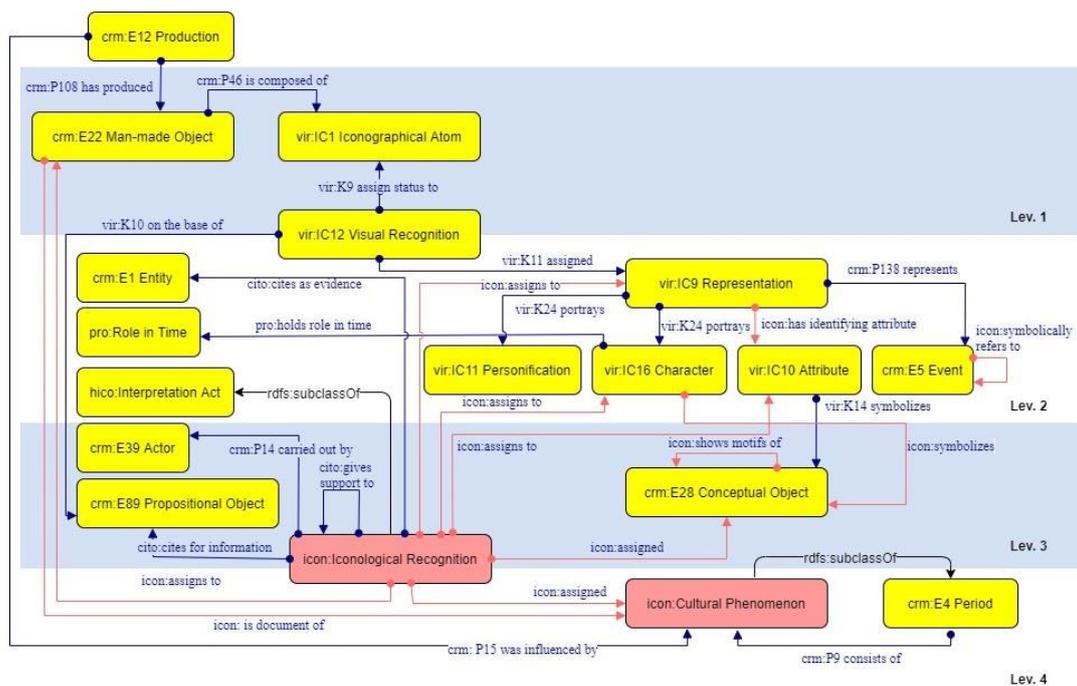

Fig. 2: Overview of classes and properties that are relevant to iconological studies

The class *vir:IC1_Iconographical_Atom* represents physical arrangements of forms and colours that will be the object of a pre-iconographical analysis (Carboni, 2019; Panofsky, 1962) (lev. 1).

---

[51] All pictures are available in Public Domain at <https://commons.wikimedia.org/>, except 1.d, published by <https://www.canalgrandevenezia.it/> under a Creative commons 3.0 license.
[52] <http://www.cidoc-crm.org/cidoc-crm/> (last consulted: 21/04/2021).
[53] <http://w3id.org/vir/> (last consulted: 21/04/2021).
[54] <http://purl.org/spar/pro/> (last consulted: 21/04/2021).
[55] <http://purl.org/emmedi/hico> (last consulted: 21/04/2021).
[56] <http://purl.org/spar/cito/> (last consulted: 21/04/2021).

From an atom, one or more representations (*vir:IC9_Representation*) can be recognized (a *vir:IC12_Visual_Recognition*) thanks to features depicted in the visual surface, such as attributes (*vir:IC10_Attribute)*, characters (*vir:IC16_Character*), personifications (*vir:IC11_Personification*), narrative aspects (*crm:E5_Event*), or concepts, e.g. allegories meaning (*crm:E28 Conceptual Object*) (lev. 2). The historian leverages artist's sources (*crm:E89 Propositional Object*) to support the interpretation (*vir:K10 on the base of*) and historical sources to collect information on represented subjects, such as historical and political roles of personifications (*pro:RoleInTime*).

The historian (*crm:E39_Actor*) performs an iconological recognition (*icon:IconologicalRecognition*) where representations, attributes, and characters are associated (*icon:assignsTo*) to meanings that the artist or the client could have consciously expressed (lev. 3). Meanings are represented by the class *crm:E28_Conceptual_Object*, that is, concepts expressed by symbols and subjects individuated in the previous level (Van Straten, 1994), related through the shortcuts *vir:K14 symbolize, icon:symbolizes* or the longer path *icon:IconologicalRecognition, icon:assignsTo*. The evidence is cited by means of *cito:citesAsEvidence*; the relation *cito:citesForInformation* individuates the possible bibliographic reference.

Finally (lev. 4), the historian relates the artwork analyzed (*icon:assignsTo*) to a cultural phenomenon (*icon:CulturalPhenomenon*) that may have influenced (*crm:P15_was_influenced_by*) the artwork production (*crm:E12_Production*). To this extent, we can classify the artwork as a document (*icon:isDocumentOf*) of the existence of the phenomenon.

*5.1. Attribution of subjects to different interpretation levels*

In Vermeer's *Woman Holding a Balance* (Fig. 1, a), Van Straten identified the standing woman as a personification of the Divine Justice (lev. 2). The recognition is based on the association of the act of weighting an empty balance (lev. 1), to the background painting depicting the Last Judgement - in which Christ weights and judges souls at the end of times according to their moral conduct. The symbolic meaning of other objects in the room supports this interpretation, namely: the mirror hung on the wall in front of the woman's face (symbol of introspection), and the jewellery box on the table (vanitas of earthly goods). The relation between the two acts of weighting, one performed with real objects and a metaphorical one in the painting, is the key for reading the work: the woman is morally weighting her conduct as an invitation for the observer to do the same (lev. 3). Assumptions on the purpose of the artwork in the context of the XVII century Dutch society (lev. 4) have been proposed. In Vermeer's *Allegory of Faith*, a similar home interior setting, and the catholic theme, have been associated with the prohibition to practice catholic faith publicly[57]. Therefore, both paintings can be deemed documents of the historical phenomenon.

In this scenario we identified the following competency questions: *What is the relation between the act of weighting an empty balance and the weighing of Souls in the Last Judgement? What is the relation between the artwork and the phenomenon of Catholicism prohibition? What is the evidence of the relation between the artwork and the cultural phenomenon?*

---

[57] W. LIEDTKE, *Johannes Vermeer | Allegory of the Catholic Faith,* in «The Metropolitan Museum of Art», 2010, URL <https://www.metmuseum.org/art/collection/search/437877> (last consulted: 18/03/2021).

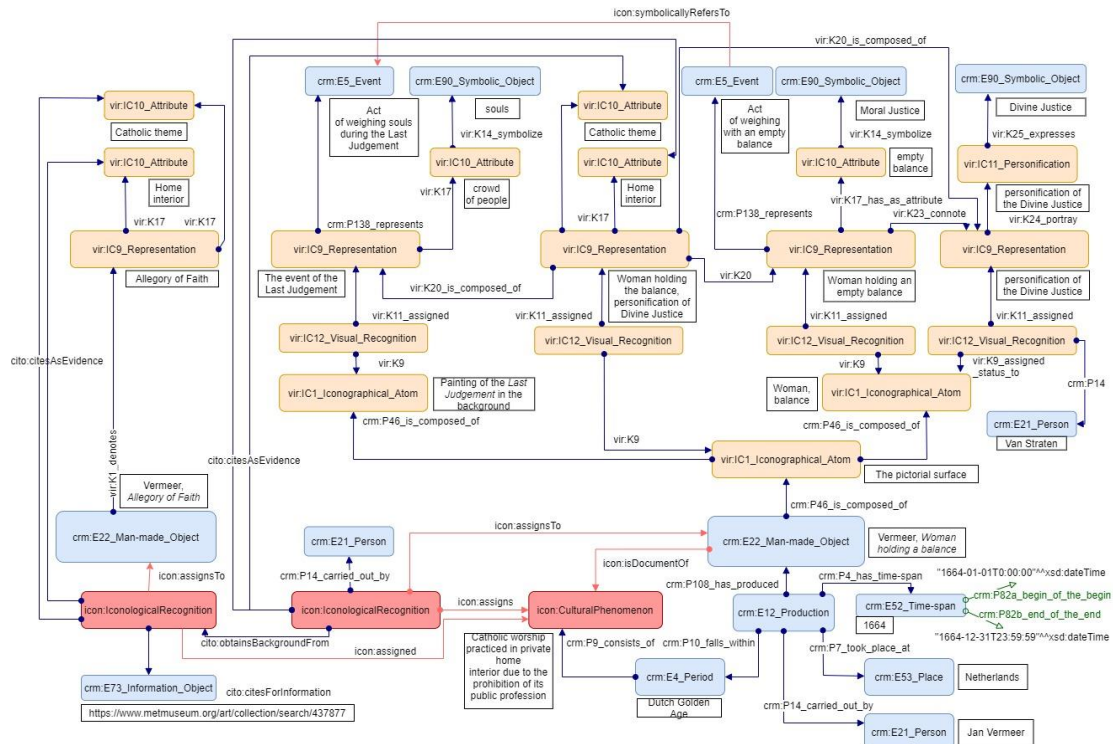

Fig. 3: Graphical representation of the Woman holding a balance

Figure 3 shows a graphical representation of the scenario. The painting presents two subjects, namely: the woman holding the balance and the painting in the background. An individual of *vir:IC1* "woman" is related to two individuals of *vir:IC9*, one representing respectively the woman holding a balance and the personification of Divine Justice. These individuals represent the same figure, but this redundance is functional to attribute only the last identification (*crm:P14*) to Van Straten. Therefore, the two representations are linked by the relation *vir:K23*, and the first is linked to the act of weighing (*crm:E5*). Likewise the representation of the Last Judgement (*vir:IC9*) is linked to the act of weighing souls (*crm:E5*).

However, the parallelism between the two actions cannot be expressed according to existing terms. For this reason, we created the property *icon:symbolicallyRefersTo* which allows us to link two individuals of the class *crm:E5*, respectively linked to the objects involved in the action. The interpretive processes are described as individuals of the class *icon:IconologicalRecognition* respectively linking the two Vermeer's paintings (*icon:assignsTo*) to the cultural phenomenon (*icon:assigns*). The analysis of *Allegory of Faith* allows us to infer a relation between the *Woman Holding a Balance* and the same cultural phenomenon. We express this dependence between the recognitions by means of the property *cito:obtainsBackgroundFrom*.

### *5.2. Diachronic changes in a subject representation*

Panofsky recognised that during the Middle Ages representations of classical themes are «expressed by nonclassical figures in a nonclassical setting»[58]. This can be explained by the fact that two traditions - textual and visual - differently transmitted classical themes from late antiquity onward, and classical themes were interpreted in a fashion closer to contemporary imagery and to the mature artistic style reached. Therefore, manuscripts carrying classical themes deprived of their classical representation allowed illuminations in which motifs reflect instead medieval representational conventions and imaginary.

---

[58] E. PANOFSKY, *Meaning in the visual arts, cit.,* p. 43

An example of this phenomenon can be found in the *Laocoon and his sons' episode* (originally taken from Vergil's *Aeneid*. 2, 199-233) represented in an illuminated medieval manuscript[59]. Here, «the Laocoon who makes the sacrifice becomes a wild and bald old priest who attacks the little bull with [...] an ax, while the two little boys float around [...] and the sea snakes appear»[60].

The scene is very different from the classical style of the Vatican statue (Fig.1.b), in which characters are semi-nude figures entangled in sea snakes. It focuses on their desperation, and on the beauty of the proportioned human body. It is worth noting that the stylistic difference between the scene depicted in the manuscript (showing medieval motifs) and the classic statue is evidence of the cultural phenomenon wherein artists revisit classic motifs according to the contemporary style.

In this case study we would like to address the following competency questions: *What are the different attributes of the subject in the two representations? What is the cultural phenomenon characterising the artists' approach?*

In Fig. 4 we represent the Vatican manuscript and the Vatican statue.

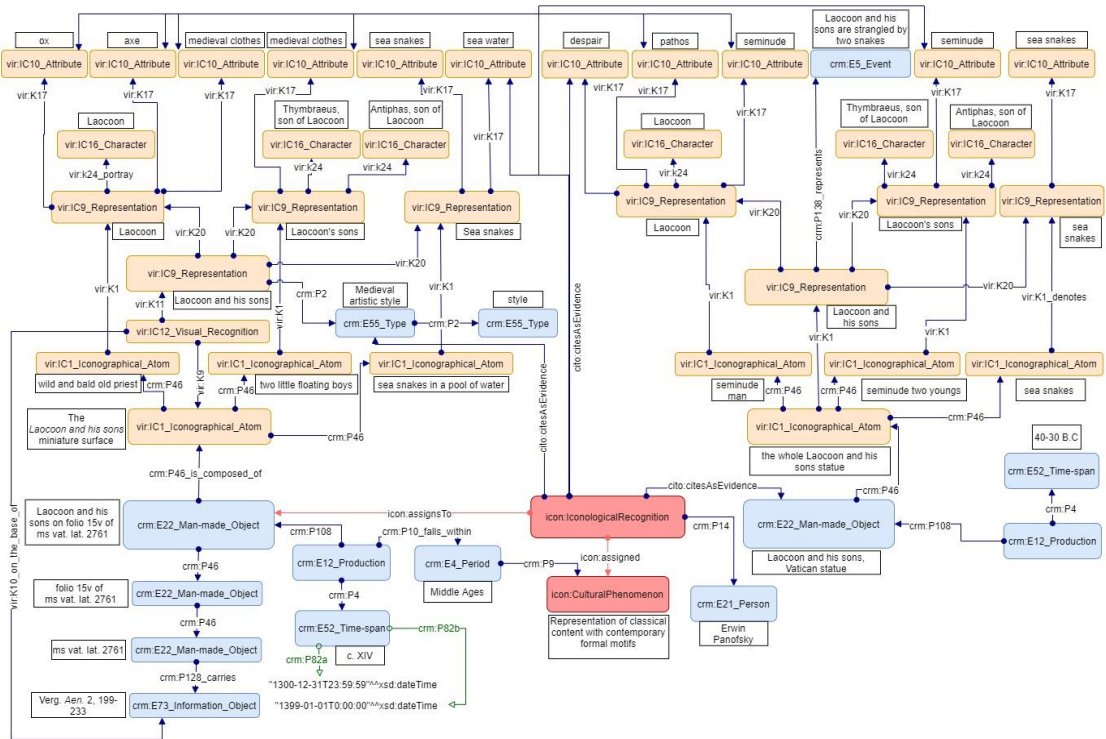

Fig. 4: Graphical representation of the Vatican manuscript depicting Laocoon and his sons and the Vatican statue

In the case study, characters (*vir:IC9*) are linked to their attributes, which show significant variants across centuries. The class *vir:IC10_Attribute* aims at describing relevant aspects of a representation that support its identification. Nevertheless, it is not possible to distinguish a relevant feature from an identifying one. We define *icon:hasIdentifyingAttribute* to relate only those attributes that have a key role in the identification of the subject to the representation at hand. We use here *vir:K17* to annotate attributes that characterize the two representations as visual evidence (*cito:citesAsEvidence*) of the iconological interpretation (lev. 4) despite not contributing to the recognition of subjects, namely: the presence of water, ox, axe and medieval clothes in the illumination, and the pathos and seminude figures in the Vatican sculpture.

---

[59] Vatican City, Biblioteca Apostolica Vaticana, Aeneis, XIV sec., ms vat. lat. 2761, folio 15r. Anonymous, *Laocoön and His Sons*. Images of the manuscript cannot be reproduced for copyright reasons. A digitization of the manuscript is available at: <https://digi.vatlib.it/view/MSS_Vat.lat.2761> (last consulted: 21/04/2021).
[60] E. PANOFSKY, F. Saxl, *op. cit.,* p. 253

## 5.3. Connections with literary sources

Irving Lavin[61] related the Neptune's statue to the urban renewal project designed by the papal vicelegate Pier Donato Cesi (1522-1586). According to the documents of the urban plan, the Neptune statue was one of the three statues designed with the aim to affirm the positivity of the papal domain over the city (lev. 3), and to remind opponents that they would be punished[62]. The iconography of the statue is influenced by Marcantonio Raimondi's decoration of the first book of Virgil's *Aeneid* of an edition commissioned by pope Julius II (Fig. 1, c). The decoration depicts *Quos Ego*, a scene in which Neptune calms a storm caused by Aeolus, therefore claiming his authority over the marine kingdom (Verg, *Aen*, I, v. 135). It's worth noting that a political message was already attributed to Virgil's text, wherein Neptune is depicted as a worthy nobleman who quells a tumult thanks to his argumentation. Lavin believes that the iconography has been reused by Cesi, ruler of Bologna, in the Neptune's statue (lev. 3)[63]. In this scenario we would like to answer the following question: *What is the symbolic meaning in common between the Neptune's statue, the iconographic source Quos Ego, and the text source of Virgil's work?*

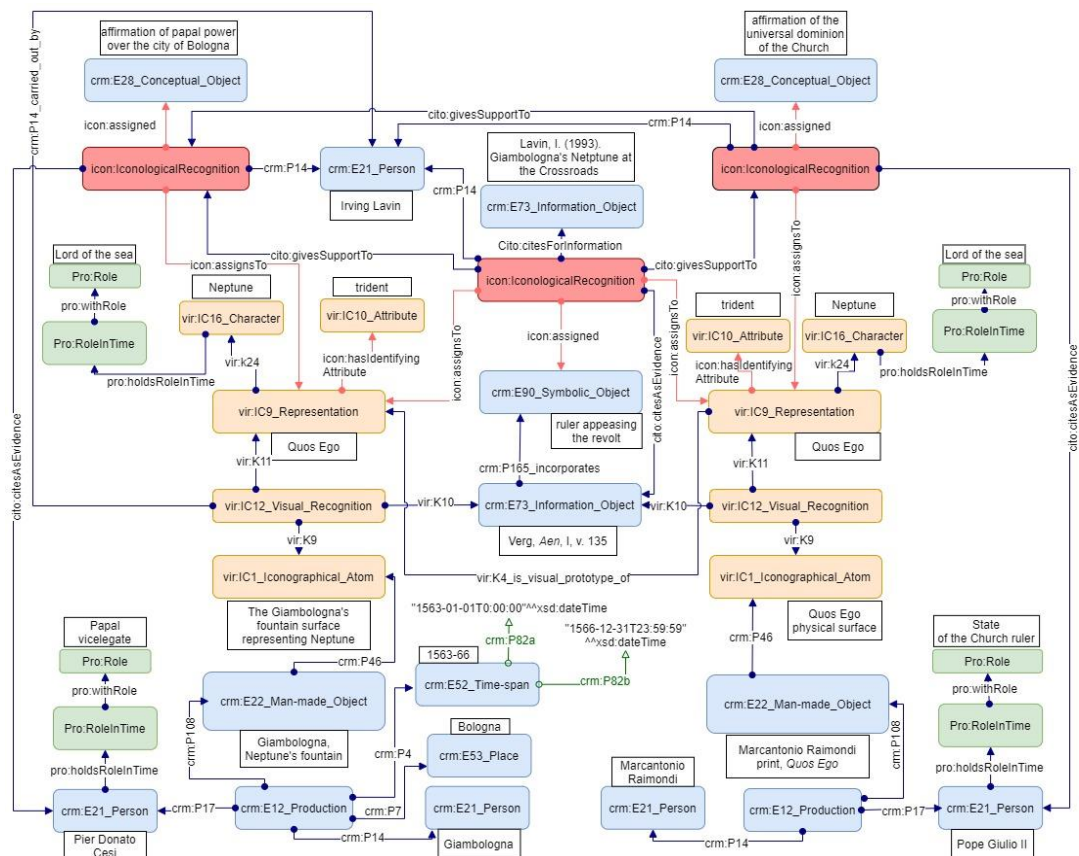

Fig. 5: Graphical representation of the Neptune's statue and the print representing *Quos Ego*.

Existing vocabularies allow us to describe subjects (*vir:IC9*) of the statue and the Quos Ego iconography (lev. 2), and their symbolic values (*crm:E90*) (lev. 3). Raimondi's representation of Quos Ego is related to the *Quos Ego* representation in Neptune's statue via *vir:K4_is_visual_prototype_of*. To highlight the fact that the former iconography has been referenced with the aim of highlighting the role of the ruler, the role of Neptune is described as an individual of *pro:Role*. Likewise, we annotate roles for pope Julius II and Pier Donato Cesi.

---

[61] I. LAVIN, *op. cit.*
[62] *Ibidem*, p. 28.
[63] *Ibidem*, pp. 75, 80.

Virgil's text is the original reference of the meaning described in both Raimondi's work and Neptune's statue and it's indicated as the source of *vir:IC12_Visual_Recognition* identifying the two artworks' representations. The meaning of ruler who quells the rioting is then related to Virgil's text by means of *crm:P165_incorporates*. An iconological recognition associates the *Quos Ego* iconography depicted in both artworks to this meaning, citing *Aeneid* as the source. Then, on the basis of such a recognition, another iconological recognition associates the *Quos Ego* represented in Raimondi's print to the concept "affirmation of the universal dominion of the Church", as it's a first christian interpretation of the meaning brought by the source. Finally, another instance of iconological interpretation supported by the previous ones relates the Giambologna's representation to the concept expressing the "affirmation of papal power over the city of Bologna".

### *5.4. Different subjects represented by the same shapes and motifs*

The case study considers the phenomenon of classical visual motifs rearranged to represent subjects representative of the Middle Ages imagery (lev. 4), as identified by Saxl and Panofsky[64]. The case study differs from the case of classical subjects revised with medieval motifs (see Section 5.2), since here the classical motifs are maintained but the conveyed theme is different. An example of this phenomenon can be found in a relief belonging to late imperial roman age representing *Hercules and the Erymanthian Boar* (Fig. 1, d first) and the medieval *Allegory of Salvation* (Fig. 1, d second) placed outside the Basilica di San Marco in Venice. Saxl and Panofsky argued that the medieval artist used the roman example as a prototype and intentionally used its motifs to represent a different subject. Despite the visual compositions being very similar, in the later work, the boar becomes a deer (lev.1), a symbol of the soul (lev. 3); king Eurystheus becomes a dragoon crushed by Christ, symbol of the devil (lev. 3); the lion skin turns into a cloth; Hercules carrying the Erymanthian boar to King Eurystheus turns into Christ who defeats the devil and saves the believers' souls[65]. Overall, the scene represents an allegory of salvation.

The competency questions we would like to answer are: *What are the attributes that allow us to recognize an iconographical subject and what are the attributes that are relevant to an iconological recognition? What is the relation between characters and attributes in the two scenes?*

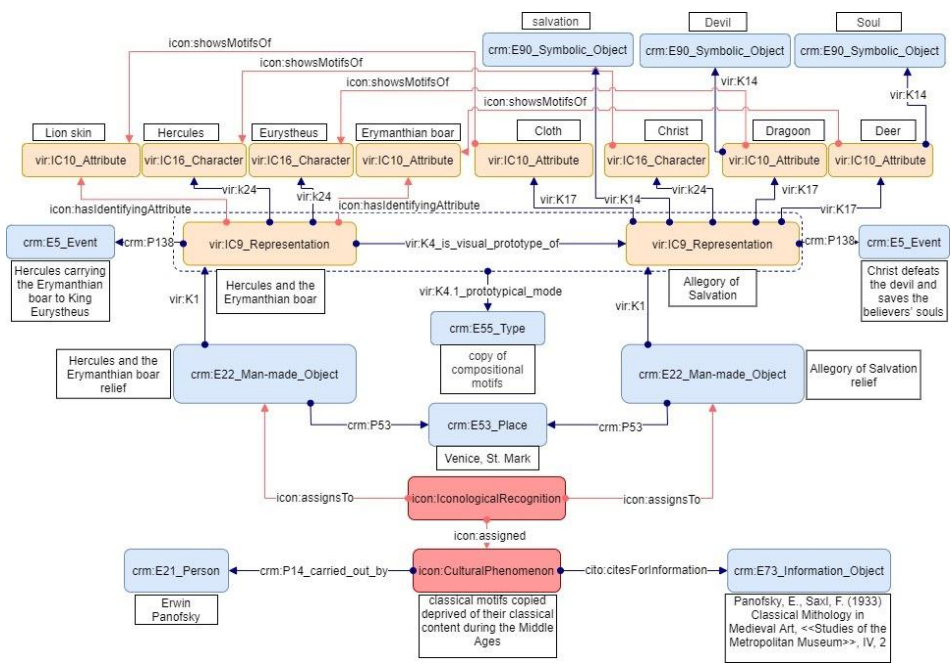

Fig. 6: Graphical representation of *Hercules and the Erymanthian Boar* and the *Allegory of Salvation*

---

[64] E. PANOFSKY, F. Saxl, *op. cit.*
[65] *Ibidem*, p. 228

All objects, subjects, and basic meanings can be represented with vocabularies, including the entire classical scene (identified as 94L3241 in Iconclass), except Eurystheus and the concept of *allegory of salvation*. By means of *crm:P53_has_current_location* we record St. Mark's Basilica as the place of both of the artworks. The property *vir:K4_has_visual_prototype* links the medieval artwork to its source and *vir:K4.1_prototipical_mode* allows to express which type of relation exists ("copy of compositional motifs") as the cultural phenomenon (lev. 4). Moreover, we add the relation *icon:showsMotifsOf* to detail the correspondence of motifs between characters and attributes of the two artworks. We use an iconological recognition to relate the two artworks with the phenomenon of classical contents reused as deprived of their original content.

## 6. Preliminary evaluation

Several methodologies for ontology evaluation exist[66]. Since a new ontology has not been developed yet, we perform a preliminary, qualitative, evaluation. Comparing ontologies to golden standards[67] is an evaluation method that allows us to demonstrate the consistency of the taxonomy and the relations by comparing them to those of existing ontologies that we define as golden standards. The aim is to demonstrate usefulness, originality, and feasibility of a future ontology dedicated to iconological studies. In Fig. 7 we provide an overview of the alignments to existing ontologies.

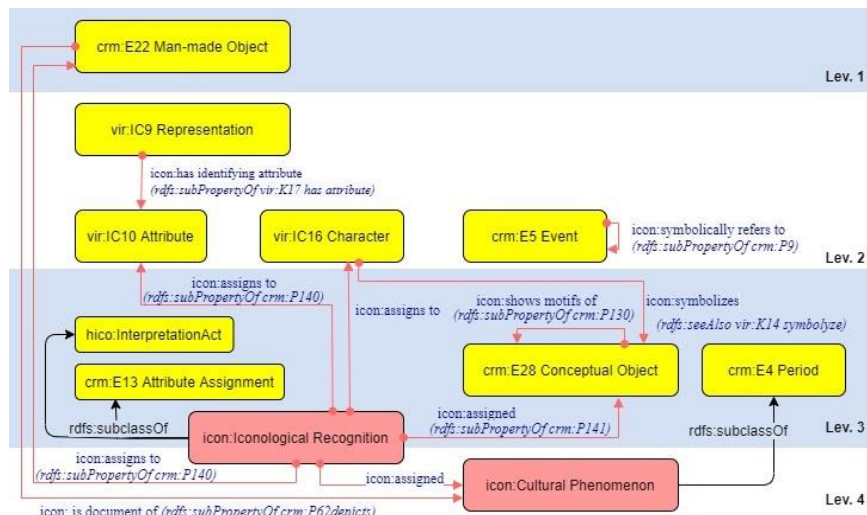

Fig. 7: Alignment to existing ontologies

We aligned icon properties and classes to CIDOC-CRM and VIR by means of RDF constructs *rdfs:subPropertyOf* and *rdfs:subClassOf*. The property *icon:hasIdentifyingAttribute* is defined as a subproperty of *vir:K17_has_attribute*. Despite attributes should be recorded when these allow to recognize the subject, in section 5.2, we addressed attributes that do not identify a subject but are evidence of cultural phenomena. Likewise, *icon:symbolicallyRefersTo* is a subproperty of *crm:P9_consists_of* that allows us to relate events. We believe the alignment is not precise, since CIDOC-CRM does not allow associative relations between spatiotemporal entities. Moreover, the relation between depicted elements is the result of a recognition, and as such it should be represented by means of the longer pattern *icon:IconologicalRecognition / icon:assigned*.

[66] J. BRANK, M. GROBELNIK, D. MLADENIC, *A survey of ontology evaluation techniques*, in "Proceedings of the conference on data mining and data warehouses (SiKDD 2005)", vol. XXVII, Citeseer Ljubljana, Slovenia October 2015.
[67] A. MAEDCHE, S. STAAB, *Measuring similarity between ontologies, in "*Proceedings of CIKM 2002", in LNAI, vol. 2473.

*icon:showsMotifsOf* is a subproperty of *crm:P130_shows_features_of* with domain and range *crm:E28_Conceptual Object*. It relates a representation (i.e. a character or an attribute) to another representation showing the same formal motifs and could be a prototype. The relation can happen between different subjects with formal similarity, as found in section 5.4. *icon:symbolizes*. We found in case study 3 (typology 1), here not described[68], that a symbolic value could be assigned to a character as well as to an attribute. We found a reference in *vir:K14_symbolize*, for which a hierarchical alignment is not possible due to the restriction on the property domain, that does not include the class *vir:IC16_Character*. The property is meant to be a shortcut of the longer pattern *icon:IconologicalRecognition / icon:assigned*, therefore a workaround is possible.

*icon:IconologicalRecognition* is defined as a subclass of *crm:E13_Attribute Assignment*, defined as an activity in which a person proposes a claim (*icon:assigned*) about an object of interest (*icon:assignsTo*). However, the relations between individuals of class *crm:E13* and the relation between these and the sources used to perform the activity are not extensively described in CIDOC-CRM. For these reasons, we also align the *icon:IconologicalRecognition* to the class *hico:InterpretationAct*. According to HiCO, an interpretation activity can be referenced by another activity (e.g. to support a new claim - via *cito:agreesWith* -, to disagree with it - via *cito:disagreesWith*) and can be annotated with the heterogeneous sources leveraged specifying their different purpose (e.g. being sources of information, evidence of a claim). Such relations are expressed by reusing the taxonomy of properties available in CiTO. *icon:CulturalPhenomenon* is defined as a subclass of *crm:E4_Period.* The concept represents a spatiotemporal entity which can be composed of (*crm:P9*) other entities. Therefore, we can create hierarchical relations between broad historical periods (e.g. the Middle Ages) and periods identifying phenomena that may cross, overlap, and depend on the former. *icon:isDocumentOf* is defined as a subproperty of *crm:P62_depicts* to relate an artefact and a cultural phenomenon. The alignment is not a perfect fit, since CIDOC-CRM addresses only visual representation as a possible relation between a physical object and an abstract, spatiotemporal, concept. Nonetheless, we believe that recording a direct relation between an object and the phenomenon that was recognized as influencing its production would be an efficient shortcut for longer patterns *icon:assigned / icon:assignedTo*.

## 7. Discussion and conclusion

From the analysis of the case studies emerged that most visual aspects relevant to iconographical interpretations can be described by means of existing ontologies and thesauri. Yet, we found a few aspects relevant to iconography and iconology are missing. We highlighted how associative relations between depicted subjects can be relevant to an iconological recognition (see 5.1), as well as the role of attributes (see 5.2, 5.4) and the representation of symbolic meanings deduced from a source (see 5.3). Moreover, we addressed how a cultural phenomenon can help the interpretation of the purpose of the artwork (5.1), and vice versa, the analysis of stylistic or formal features can disclose underlying cultural phenomena (5.2, 5.4). Furthermore, several types of relations between historians' claims, and between claims and their sources, should be addressed.

In the preliminary evaluation, we were able to align almost all terms to existing ontologies, demonstrating that the consistency of the proposed taxonomy and relations would be inherited from golden standards. We also discussed some inaccuracies in the alignment, which provide us with a genuine motivation for developing a bespoke ontology in the future that extends CIDOC-CRM, VIR, and HiCO. For instance, a class representing iconological interpretations allows us to specify the evidence and the related interpretations supporting a claim, and to group different artworks linked to the same cultural phenomenon. Future works include the creation of a dataset relevant to iconological studies as a proof of concept, and the evaluation of the ontology on real-world data.

---

[68] The case study is not presented in the article due to template limits. A detailed description is available at <https://sofibar.github.io/Icon/>.

Lastly, we envision an application-based evaluation of the ontology, wherein to define art historical patterns in an ontology-driven fashion and to show benefits in terms of knowledge discovery.

## References

ALEXIEV, V., *Museum linked open data: Ontologies, datasets, projects*, In «Digital Presentation and Preservation of Cultural and Scientific Heritage», vol. VIII, 2018, pp. 19-50.

BRANK, J., GROBELNIK, M., MLADENIC, D., *A survey of ontology evaluation techniques*, in "Proceedings of the conference on data mining and data warehouses (SiKDD 2005)", vol. XXVII, Citeseer Ljubljana, Slovenia, October 2015.

CARBONI, N., DE LUCA, L., *An Ontological Approach to the Description of Visual and Iconographical Representations*, in «Heritage», vol. II, n. 2, 2019, pp. 1191–1210, doi: https://doi.org/10.3390/heritage2020078.

CARDINI, F., ACIDINI LUCHINAT, C., RICCIARDI, L., *I re magi di Benozzo a palazzo Medici*. Mandragora, Firenze 2001

CARRIERO, V., et al., *The landscape of ontology reuse approaches*, in «Applications and Practices in Ontology Design, Extraction, and Reasoning», vol. XLIX, n. 21, 2020.

CARRIERO, V. A., GANGEMI, A., MANCINELLI, M. et. al., *ArCo: The Italian cultural heritage knowledge graph*, in *International Semantic Web Conference*, edited by Springer, Cham, October 2019, pp. 36-52.

CHRISTIANSEN, K., *Lorenzo Lotto and the Tradition of Epithalamic Painting*, in «Apollo», n.124, 1986, pp. 166–73.

CHRISTIANSEN, K., *Lorenzo Lotto | Venus and Cupid*, in «The Metropolitan Museum of Art», 2018, URL <https://www.metmuseum.org/art/collection/search/436918>.

DAQUINO, M., MAMBELLI, F., PERONI, S. et. al., *Enhancing semantic expressivity in the cultural heritage domain: exposing the Zeri Photo Archive as Linked Open Data*, in «Journal on Computing and Cultural Heritage (JOCCH)», vol. X, n. 4, 2017, pp. 1-21.

DAQUINO, M., TOMASI, F., *Historical Context Ontology (HiCO): a conceptual model for describing context information of cultural heritage objects*, in *Research Conference on Metadata and Semantics Research*, edited by Springer, Cham, September 2015, pp. 424-436.

DIJKSHOORN, C., JONGMA, L., AROYO, L. et. al., *The Rijksmuseum collection as linked data*, in «Semantic Web», vol. 9, n.2, 2018, pp. 221-230.

DOERR, M., GRADMANN, S., HENNICKE, S. et. al., *The europeana data model (edm).*, in «World Library and Information Congress: 76th IFLA general conference and assembly», vol. X, August 2010, p. 15.

DOERR, M., *The CIDOC conceptual reference module: an ontological approach to semantic interoperability of metadata,* in «AI magazine», vol. XXIV, n. 3, 2003, 75-75.

GARTNER, R., *Towards an ontology-based iconography*, in «Digital Scholarship in the Humanities», vol. XXXV, n. 1, 2020, pp. 43-53.

GOMBRICH, E. H., *Symbolic images*, Phaidon, London 1972.

HIRN, Y., *The Sacred Shrine: A study of the poetry and art of the catholic church*, Beacon Press, Boston 1957, pp. 296-8.

IMDAHL, M., *Iconica. L'intuizione delle immagini*, in «Aisthesis. Pratiche, Linguaggi e Saperi dell'estetico», vol. v, n. 2, 2012, pp. 11-32. doi: https://doi.org/10.13128/Aisthesis-11474.

KNOBLOCK, C. A., SZEKELY, P., FINK, E. et. al., *Lessons learned in building linked data for the American art collaborative*, *International Semantic Web Conference*, edited by Springer, Cham, October 2017, pp.263-279.

LAVIN, I., *Giambologna's Neptune at the Crossroads*, in LAVIN, I., *Past-present: essays on historicism in art from Donatello to Picasso*, University of California Press, Berkeley 1993, pp. 63-84.

W. LIEDTKE, *Johannes Vermeer | Allegory of the Catholic Faith,* in «The Metropolitan Museum of Art», 2010, URL <https://www.metmuseum.org/art/collection/search/437877> (accessed 3.18.21).

MAEDCHE, A., STAAB, S., *Measuring similarity between ontologies, in "*Proceedings of CIKM 2002", LNAI, vol. 2473.

L. MANOVICH, *Data science and digital art history*, in «International Journal for Digital Art History», vol. I, 2015.

MITCHELL, W. J. T., *Iconology: image, text, ideology*, The University of Chicago Press, Chicago 2013.

MÜLLER, M. G., *Iconography and Iconology as a Visual Method and Approach*, in MARGOLIS, E., PAUWELS, L. (ed.) *The SAGE Handbook of Visual Research Methods*, SAGE Publications Ltd, 2011, pp. 283–97.

PANOFSKY, E., *Studies in iconology: Humanistic themes in the art of the Renaissance*, Harper & Row, New York 1962.

PANOFSKY, E., *The iconography of Correggio's camera di San Paolo,* The Warburg Institute University of London, London 1961.

PANOFSKY, E., *Meaning in the visual arts: papers in and on Art History,* Doubleday, Garden City, N.Y 1955.

PANOFSKY, E., SAXL, F., *Classical Mythology in Medieval Art*, in «Metropolitan Museum Studies», vol. IV, n. 2, 1933, pp. 228-280.

PERONI, S., SHOTTON, D., *The SPAR ontologies*, in *International Semantic Web Conference*, edited by Springer, Cham, October 2018, pp. 119-136.

PINELLI, O. R., *La Storia delle storie dell'arte*, Einaudi, Torino 2014.

ROBB, D. M., *The iconography of the Annunciation in the Fourteenth and Fifteenth Centuries*, in «The Art Bulletin», vol. XVIII, n. 4, 1936, pp. 480-526.

VAN STRATEN, R., *An Introduction to Iconography: Symbols, Allusions and Meaning in the Visual Arts*, Taylor & Francis, New York 1994.
WARBURG, A., *The renewal of pagan antiquity. Contributions to the cultural history of the European Renaissance*, Getty Research Institute for the History of Art and the Humanities, Los Angeles 1999.
WITTKOWER, R., *Allegory and the migration of symbols,* Thames and Hudson, London 1987.
WITTKOWER, R*., Chance, Time and Virtue*, in «Journal of the Warburg Institute», vol. I, n. 4, 1938, pp. 313-321. doi:10.2307/749998.